\documentclass[preprint,12pt,authoryear]{elsarticle}

\usepackage{amssymb}
\usepackage{amsmath}
\usepackage{amssymb}
\usepackage{changepage}
\usepackage{float}
\usepackage{dsfont}
\usepackage{natbib}
\usepackage{graphicx}
\usepackage{amsmath}
\usepackage{graphics}
\usepackage{multirow}
\usepackage{rotating}
\usepackage{lscape}
\usepackage{graphicx}
\usepackage{subcaption}
\usepackage {comment}
\usepackage [normalem] { ulem }
\usepackage {soul}
\usepackage{xcolor}
\usepackage{graphicx}
\usepackage{float} 
\usepackage{hyperref}
\usepackage{tikz}
\usetikzlibrary{arrows.meta, positioning}
\DeclareMathOperator{\diag}{diag}

\journal{Spatial statistics}

\begin{document}

\begin{frontmatter}



\title{Efficient bayesian spatially varying coefficients modeling for censored data using the vecchia approximation}
\author[label1]{Idir Mohamed Yacine}
\author[label1]{Romary Thomas}

\affiliation[label1]{organization={Ecole des mines de Paris}, 
            addressline={35 Rue Saint-Honoré}, 
            city={Fontainebleau},
            postcode={77300}, 
            state={Ile de france}, 
            country={France}}

\begin{abstract}

Spatially varying coefficients (SVC) models allow for marginal effects to be non-stationary over space and thus offer a higher degree of flexibility with respect to standard geostatistical models with external drift. At the same time, SVC models have the advantage that they are easily interpretable. They offer a flexible framework for understanding how the relationships between dependent and independent variables vary across space.

The most common methods for modelling such data are the Geographically Weighted Regression (GWR) and Bayesian Gaussian Process (Bayes-GP).
The Bayesian SVC model, which assumes that the coefficients follow Gaussian processes, provides a rigorous approach to account for spatial non-stationarity. However, the computational cost of Bayes-GP models can be prohibitively high when dealing with large datasets or/and when using a large number of covariates, due to the repeated inversion of dense covariance matrices required at each Markov chain Monte Carlo (MCMC) iteration.
In this study, we propose an efficient Bayes-GP modeling framework leveraging the Vecchia approximation to reduce computational complexity while maintaining accuracy. The proposed method is applied to a challenging soil pollution data set in Toulouse, France, characterized by a high degree of censorship (two-thirds censored observations) and spatial clustering. 

Our results demonstrate the ability of the Vecchia-based Bayes-GP model to capture spatially varying effects and provide meaningful insights into spatial heterogeneity, even under the constraints of censored data.

\end{abstract}


\end{frontmatter}



\newpage

\section{Introduction}
Soil pollution by various contaminants, including hydrocarbons, heavy metals and pesticides, poses significant dangers to both environmental and human health. These contaminants are highly persistent in the environment, bioaccumulate in food chains, and can be transported across different environmental compartments including groundwater, ultimately reaching humans through direct contact, inhalation of contaminated dust, or consumption of polluted food and water \citep{mekonnen2024bioremediation,eea_soil_pollution}

Exposure to hydrocarbon-polluted soils is linked to a range of adverse health effects, including genetic mutations, immunotoxicity, neurotoxicity, teratogenicity, and an increased cancer risk due to the carcinogenic nature of many hydrocarbon compounds \citep{venkatraman2024environmental}. Recent reviews highlight that soil pollution, often less visible than air pollution, contributes to cardiovascular diseases, respiratory illness, and other chronic health conditions through mechanisms like oxidative stress and inflammation \citep{munzel2025soil}. Given these substantial health risks, it is crucial that public agencies are provided with accurate, interpretable, uncertainty-aware information when assessing and managing soil hydrocarbon contamination. 


The factors influencing hydrocarbon distribution in soils operate across multiple scales, creating complex spatial patterns that challenge accurate mapping. At the microscale, soil physicochemical properties including organic matter content, clay percentage, and pH strongly influence hydrocarbon retention and degradation rates \citep{shu2024spatial}. Mesoscale factors include hydrological processes, particularly in areas with shallow groundwater where contaminant transport can create dynamic pollution plumes \citep{lee2001factors}. At the landscape scale, anthropogenic activities, especially industrial operations, transportation networks, and waste disposal practices, create distinctive spatial patterns that require specialized sampling strategies \citep{froger2021spatial}. 

Accounting for cross-scale interactions to effectively map soil pollution is particularly challenging, especially in areas with sparse sampling. To compensate for the lack of direct pollutant measurements, covariates, available across the entire area, are commonly used. These covariates help enrich the data and, in some cases, reveal hidden relationships between hydrocarbon distribution and the covariates themselves. Various models incorporate such auxiliary information with varying levels of interpretability, from kriging with external drift  \citep{de2006modelisation}, to machine learning approaches \citep{meng2023shaping}.

Spatially Varying Coefficients (SVC) has regained recent interest \citep{comber2024multiscale,finley2020bayesian,mu2018estimation}, as it offers a nice middle ground between the two worlds, by letting the relationships between dependent and independent variables vary across space, it offers great flexibility and at the same time retains all the interpretability \citep{finley2011comparing}.

The most common methods for modeling such data are the Geograph-
ically Weighted Regression (GWR) \citep{fotheringham2009geographically} and  Gaussian Process SVC (GP-SVC) \citep{dambon2021maximum}. The GP-SVC model, which assumes that the coefficients follow
Gaussian processes, provides a rigorous approach to account for spatial non-stationarity. However, the computational cost of GP-SVC models can be
prohibitively high when dealing with large datasets or/and when using a
large number of covariates, due to the need for repeated inversion of dense
covariance matrices.\\

Uncertainty quantification using hierarchical Bayesian models has become a key component of modern soil pollution mapping, offering a comprehensive framework for risk assessment by jointly addressing measurement error, model parameter uncertainty, and spatial interpolation variability. This multi-level approach provides more reliable uncertainty estimates than single-scale methods, especially in areas with sparse sampling \citep{banerjee2003hierarchical}.

Bayesian spatially varying coefficient models are powerful too ls due to their flexibility. However, this increased flexibility comes at a computational cost: "By far the greatest challenge to widespread adoption of these models is computational." concluded \citep{finley2011comparing}. 
While recent advances in the frequentist context have significantly improved scalability \citep{dambon2021maximum, MURAKAMI201939}, the Bayesian framework—where uncertainty quantification is most natural—remains hampered by prohibitive computational demands. Only a handful of studies have begun to tackle this bottleneck. Notable examples include \citep{finley2020bayesian}, which implements parallelization and other computational enhancements to improve efficiency, and the more recent work of \citep{lei2024scalable}, which introduces a scalable framework based on low-rank decomposition techniques.\\

Since the problem comes from inverting large matrices, Vecchia approximation \citep{vecchia1988estimation}, which was originally proposed to approximate the likelihood, can help to solve this problem by inverting matrices of maximum size $M\ll N$ \citep{datta2016hierarchical}.
This approximation provides substantial computational gains while maintaining a high level of accuracy \citep{katzfuss2021general}. The application of Vecchia's approximation introduces some choices to be made, apart from the choice of M (the number of conditioning neighbors), it is necessary to order the observations before \citep{guinness2018permutation} as well as to choose whether to apply this to the observations directly or to the latent process \citep{finley2019efficient}.\\

Measurements in the real world often come from instruments with bounded scales or from events observed over finite time periods. As a result, the problem of censored data arises: we only observe values up to a certain threshold.

Frequentist models for censored data mostly consider the popular EM algorithm \citep{dempster1977maximum} to obtain estimates of model parameters and a variation of the Tobit model \citep{basson2023variational} in the Bayesian framework.

The Tobit model has since become a cornerstone of censored modeling in multiple research fields, including econometrics and survival analysis \citep{bewick2004statistics}. It has been extended through multiple variations \citep{greene2005censored}, such as: multiple latent variables, e.g., one variable determines which observations are non-censored while another determines their values \citep{amemiya1984tobit}; Tobit-like models of censored count data \citep{terza1985tobit}; Tobit Quantile Regression \citep{powell1986censored}; dynamic, autoregressive Tobit \citep{wei1999bayesian}; and combination with Kalman Filter \citep{allik2015tobit}. 
As reflected in all the aforementioned works, research on censored modeling commonly decomposes the process into two components: a latent process representing the true underlying values, and an observation mechanism that yields a distorted version of this latent process.

While the introduction of a latent process is perfectly interpretable and naturally aligns with real-world constraints, such as sensor limitations, it only exacerbates an already significant computational burden in bayesian GP-SVC models.
For instance, \citep{finley2019efficient}, who applied the Vecchia likelihood to a Bayesian Gaussian process (in a setting involving significantly fewer parameters than a Bayesian GP SVC model), noted:  "We observed that, for very large spatial datasets, sequential updating of the random effects often leads to very poor mixing in the MCMC. The computational gains (using Vecchia) per MCMC iteration is thus offset by a slow converging MCMC"
 \citep{liu1994covariance} showed that MCMC algorithms where one or more variables are marginalized out
tend to have lower autocorrelation and improved convergence behavior.\\

In summary, when dealing with difficult, sparse, and clustered data such as hydrocarbon contamination or soil pollution, we need a flexible model that can handle uncertainty and, crucially, remains interpretable. Because data is limited, it becomes essential to identify and incorporate relevant covariates. Therefore, the model must be able to handle covariates effectively and transparently. Bayesian GP-SVC models meet all these criteria, although they come with a high computational cost when N and/or p is large.

Moreover, when working with censored data, the modeling task becomes even more challenging. While the Bayesian framework naturally accommodates censoring through latent variables, this further amplifies issues of identifiability and computational burden.

This paper aims to address both of these challenges—censored data and computational cost—by proposing a Vecchia approximation of the Bayesian GP-SVC model that can handle censoring without introducing a latent field.

The paper is organized as follows: Section 2 introduces the proposed model after a brief review of Bayesian SVC models. Section 3 validates the proposed approximation by means of a comparative analysis. Section 4 presents an illustrative case study on hydrocarbon soil pollution in Toulouse, France. The paper concludes with a summary of the findings and perspectives for future work.

\section{Model}
Let $Z(s_i)$ and $x(s_i)$ denote the response and the predictors observed at location $s_i$, $i = 1, 2, \ldots, n$. The spatial location $s_i \in D \subset \mathbb{R}^d$, $d \geq 1$, lies within a domain $D$. In this work, we focus on the two-dimensional case ($d = 2$).

This section outlines the logical development of a Bayesian SVC model tailored to censored data, following a brief reintroduction of the standard SVC formulation. We then describe two alternative strategies for incorporating the Vecchia approximation into this framework.

\subsection{Review of Bayesian Spatially Varying Coefficients}
\subsubsection{Spatially Varying Coefficients (SVC) Model}
The SVC framework extends the standard linear regression model with $p$ predictors:

\begin{equation}
Z_i = \beta_1 x_i^{(1)} + \cdots + \beta_p x_i^{(p)} + \varepsilon_i, \quad \varepsilon_i \overset{\text{iid}}{\sim} \mathcal{N}(0, \tau^2),
\end{equation}

When the coefficients are allowed to vary across space, the model is expressed as:

\begin{equation}
Z_i = \beta_1(s_i) x_i^{(1)} + \cdots + \beta_p(s_i) x_i^{(p)} + \varepsilon_i, \quad \varepsilon_i \overset{\text{iid}}{\sim} \mathcal{N}(0, \tau^2),
\end{equation}

Following \citep{gelfand2003spatial} and more recently \citep{dambon2021maximum}, each spatially varying coefficient $\beta_j(\cdot)$ is modeled by a Gaussian process (GP). 

For each covariate $j$, the associated coefficient $\beta_j(\cdot)$ is decomposed into a fixed and a random effect:
\begin{equation} 
\beta_j(\cdot) = \alpha_j + \eta_j(\cdot)
\end{equation}

The fixed effect $\alpha_j$ represents the global mean of the coefficient, and $\eta_j(\cdot)$ is a zero-mean Gaussian process with covariance matrix $\Sigma$ derived from a stationary covariance function parametrized by $\Theta_j$: 
\begin{equation}
\eta_j(\cdot) \sim \mathcal{N}_n(0_n, \Sigma_{\Theta_j}).
\end{equation}

Assuming mutual prior independence among the Gaussian processes $\eta_j$, the joint random effect vector
\begin{equation}
\eta = (\eta_1^\top, \ldots, \eta_p^\top)^\top \in \mathbb{R}^{np}
\end{equation}

follows the distribution $\eta \sim \mathcal{N}_{np}(0_{np}, \Sigma_{\eta})$ with block-diagonal covariance matrix
\begin{equation}
\Sigma_{\eta} := \text{diag}(\Sigma(\Theta_1), \ldots, \Sigma(\Theta_p)).
\end{equation}

Let $X \in \mathbb{R}^{n \times p}$ denote the design matrix with elements $(X)_{ij} = x_i^{(j)}$, representing the $i$th observation of the $j$th covariate. The fixed effect component is given by $X\alpha$, where $\alpha = (\alpha_1, \ldots, \alpha_p)^\top \in \mathbb{R}^p$.

Further, we define $W \in \mathbb{R}^{n \times (np)}$ as a sparse matrix structured as:

\begin{equation}
W := \big(\text{diag}(x^{(1)}) \, | \ldots | \, \text{diag}(x^{(p)}) \big).
\end{equation}

Using this notation, the random effect component is expressed as $W\eta$. The error term $\varepsilon \sim \mathcal{N}_n(0_n, \tau^2 I_n)$ is assumed independent of $\eta$. In summary, writing the response as an $n$-dimensional vector $Z$, we obtain the GP-based SVC model:
\begin{equation}
Z = X\alpha + W\eta + \varepsilon.
\end{equation}

This results in a compact Gaussian process formulation: 
\begin{equation}
Z \sim \mathcal{N}_n(X\alpha, \, W\Sigma_{\eta}W^\top + \tau^2 I_n).
\label{freq_svc}
\end{equation}
Note that the dependence of $\Sigma_{\eta}$ on the covariance parameters $\Theta = (\Theta_1, \ldots, \Theta_p)$ has been suppressed for clarity.

\subsubsection{Bayesian SVC}
By assigning independent priors to the parameters in the model defined in Equation~\eqref{freq_svc}, we obtain the following Bayesian formulation:

\begin{equation} \label{eq:model_priors}
\begin{aligned}
Z \mid \alpha, \Theta, \tau^2 
&\sim
\mathcal{N}_n\bigl( X\alpha,\; W\,\Sigma_{\eta}(\Theta)\,W^\top + \tau^2 I_n\bigr), \\
\alpha, \Theta, \tau^2 
&\sim p(\alpha)\, p(\Theta)\, p(\tau^2).
\end{aligned}
\end{equation}

Posterior inference for the parameters can be performed using standard Markov Chain Monte Carlo (MCMC) methods \citep{robert1999monte}, since the model remains a Gaussian process with a structured covariance matrix.

However, in practice, for large numbers of observations ($n$) and/or predictors ($p$), the computational burden becomes substantial. This is primarily due to the need to repeatedly compute and invert the $n \times n$ covariance matrix at each MCMC iteration. The problem is further exacerbated by the high-dimensional parameter space, as each coefficient-specific Gaussian process introduces its own set of covariance parameters. This contrasts with classical GP models, which typically involve only a small number of hyperparameters.

\subsection{Bayesian SVC for censored data}

Beyond computational challenges, the presence of left-censored observations—where the values fall below a known detection threshold $L$—introduces additional methodological complexities. A principled and interpretable approach introduces a latent spatial process representing the unobserved true contamination field, with the censoring interpreted as a limitation of the measurement system. This strategy, hereafter referred to as the *full latent approach*, aligns well with empirical realities and has been shown to yield superior performance, as will be illustrated in subsequent sections.

\subsubsection{Full latent formulation}

We first consider the case where all data are fully observed. Let $w_i$ denote the underlying spatial signal at location $s_i$, defined as
\begin{equation}
w_i = \beta_1(s_i) x_i^{(1)} + \cdots + \beta_p(s_i) x_i^{(p)}.
\end{equation}
The observed response is then modeled as
\begin{equation}
Z_i = w_i + \varepsilon_i, \quad \varepsilon_i \overset{\text{iid}}{\sim} \mathcal{N}(0, \tau^2).
\end{equation}
This latent formulation is equivalent to the Bayesian hierarchical model in Equation~\eqref{eq:model_priors}, and can be rewritten as:
\begin{equation} \label{eq:model_priors_latent}
\begin{aligned}
Z \mid w, \tau^2 &\sim \mathcal{N}_n(w, \tau^2 I_n), \\
w \mid \alpha, \Theta &\sim \mathcal{N}_n\bigl( X\alpha,\; W\,\Sigma_{\eta}(\Theta)\,W^\top \bigr), \\
\alpha, \Theta, \tau^2 &\sim p(\alpha)\, p(\Theta)\, p(\tau^2).
\end{aligned}
\end{equation}
Here, $w \in \mathbb{R}^n$ represents the unobserved true latent process, while $Z$ are noisy measurements of this process, corrupted by additive Gaussian noise with variance $\tau^2$.

In the presence of left-censored data, only partial information is available about some elements of $Z$. Let $L \in \mathbb{R}$ be the known detection threshold, and define the index sets
\begin{equation}
\mathcal{I}_o = \{i : w_i + \varepsilon_i > L\}, \qquad 
\mathcal{I}_c = \{i : w_i + \varepsilon_i \leq L\},
\end{equation}
corresponding respectively to the observed and censored observations. The data then satisfy
\begin{equation}
Z_i =
\begin{cases}
w_i + \varepsilon_i, & i \in \mathcal{I}_o, \\
L,                  & i \in \mathcal{I}_c.
\end{cases}
\end{equation}
We define the partition $Z = Z_o \cup Z_c$, with
\[
Z_o = \{Z_i : i \in \mathcal{I}_o\}, \qquad Z_c = \{Z_i : i \in \mathcal{I}_c\}.
\]

The corresponding censored-data likelihood leads to the following extended model:
\begin{equation} \label{eq:model_priors_latent_censored}
\begin{aligned}
Z_i \mid w_i, \tau^2 &\sim
\begin{cases}
\mathcal{N}(w_i, \tau^2), & i \in \mathcal{I}_o, \\[0.5em]
\Pr(Z_i \leq L \mid w_i, \tau^2) = \Phi\left( \dfrac{L - w_i}{\tau} \right), & i \in \mathcal{I}_c,
\end{cases} \\
w \mid \alpha, \Theta &\sim \mathcal{N}_n\bigl( X\alpha,\; W\,\Sigma_{\eta}(\Theta)\,W^\top \bigr), \\
\alpha, \Theta, \tau^2 &\sim p(\alpha)\, p(\Theta)\, p(\tau^2),
\end{aligned}
\end{equation}
where $\Phi(\cdot)$ denotes the cumulative distribution function of the standard normal distribution.

\smallskip

While this formulation provides a coherent Bayesian treatment of censored spatially-varying coefficient models, it introduces significant computational burdens. The latent vector $w$ must be treated as an additional parameter of dimension $n$, which substantially increases the parameter space and leads to slow mixing and strong autocorrelation in MCMC samplers. These challenges are well-documented in the literature, see e.g., \citep{finley2019efficient, liu1994covariance}. In particular, two main computational bottlenecks arise: (i) the high dimensionality of the latent field $w$, and (ii) the repeated inversion of dense $n \times n$ covariance matrices needed to evaluate the full conditional distributions.

\subsubsection{Latent Vecchia approximation}

To address the computational bottleneck caused by the repeated inversion of the $n \times n$ covariance matrix, we propose applying the Vecchia approximation to the latent spatial process $w$.

The Vecchia approximation is based on rewriting the joint density of the latent vector $w$ as a product of conditional densities:
\begin{equation}
p(w) = p(w_1)\prod_{i=2}^{n} p(w_i \mid w_1, \ldots, w_{i-1}) = p(w_1)\prod_{i=2}^{n} p(w_i \mid w_{v(i)}),
\label{true_like}
\end{equation}
where $v(i) = \{1, \ldots, i-1\}$ is the full conditioning set for the $i$th variable.

The key idea behind the Vecchia approximation is to replace the full conditioning set $v(i)$ with a smaller subset $\widetilde{v}(i)$ of size at most $M \ll i-1$, often chosen to correspond to spatial neighbors of observation $i$. This yields the following approximation to the joint density:
\begin{equation}
\widetilde{p}(w) = p(w_1) \prod_{i=2}^{n} p(w_i \mid w_{\widetilde{v}(i)}),
\label{approx_like}
\end{equation}
which still defines a valid multivariate normal distribution.

Under this approximation, the latent vector $w$ follows:
\begin{equation}
\widetilde{p}(w) = \mathcal{N}(X\alpha,\; (BFB^\top)^{-1}),
\end{equation}
where $F$ is a diagonal matrix and $B$ is an upper-triangular matrix (with structure determined by the conditioning sets $\widetilde{v}(i)$). The explicit construction of $F$ and $B$ is detailed in \citep{datta2016hierarchical} and \citep{katzfuss2020vecchia}, and depends on the original covariance structure $\Sigma = W \Sigma_{\eta}(\Theta) W^\top$.

Unlike the full joint density $p(w)$, the approximated density $\widetilde{p}(w)$ depends on the ordering of the observations as well as the size $M$ of the conditioning sets. If the full conditioning is used (i.e., $\widetilde{v}(i) = v(i)$), then the approximation becomes exact. However, by selecting a small number of conditioning neighbors, the computational cost can be dramatically reduced.

In particular, this approximation avoids the need to invert a full $n \times n$ covariance matrix at each MCMC iteration. Instead, it only requires inversion of matrices of size $M \times M$. Moreover, the use of conditional distributions in the Vecchia formulation facilitates parallel computation, making this approach highly scalable for large spatial datasets.

\subsubsection{Latent-Free Vecchia approximation}

We propose a novel model that leverages the Vecchia approximation not only to mitigate the computational cost of matrix inversion but also to eliminate the need for the latent spatial field $w$. This latent-free formulation allows for scalable Bayesian SVC modeling in the presence of censored data and high-dimensional predictors, even when the number of observations $n$ is large.

The likelihood of the observed and censored data $Z$ can be written as:
\begin{equation}
    p(Z) = p(Z_1)\prod_{i \in \mathcal{I}_o} p(Z_i \mid Z_{v(i)}) \prod_{i \in \mathcal{I}_c} \int_{-\infty}^{L} p(x \mid Z_{v(i)})\,dx,
\end{equation}
where $v(i)$ denotes the conditioning set for observation $i$, and the censored indices $\mathcal{I}_c$ correspond to values below the detection limit $L$.

We apply the Vecchia approximation directly to the observed vector $Z$, by replacing each conditioning set $v(i)$ with a reduced neighbor set $\tilde{v}(i)$, constrained to include only previously ordered non-censored observations. That is, for each $i$, we require:
\[
\tilde{v}(i) \subset \{j \in \mathcal{I}_o : j < i\}.
\]
This yields the approximate likelihood:
\begin{equation}
    p(Z) \approx \tilde{p}(Z) = p(Z_1)\prod_{i \in \mathcal{I}_o} p(Z_i \mid Z_{\tilde{v}(i)}) \prod_{i \in \mathcal{I}_c} \int_{-\infty}^{L} p(x \mid Z_{\tilde{v}(i)})\,dx.
\end{equation}

Each term in this expression corresponds to a Gaussian or truncated Gaussian:
\begin{equation}
\tilde{p}(Z) = 
\mathcal{N}(Z_1 \mid X_1 \alpha, F_{11}) 
\prod_{i \in \mathcal{I}_o} 
\mathcal{N}(Z_i \mid X_i \alpha + \mu_i, F_{ii}) 
\prod_{i \in \mathcal{I}_c} 
\Phi\left(\frac{L - (X_i \alpha + \mu_i)}{\sqrt{F_{ii}}}\right),
\end{equation}
where $F_{ii}$ and $\mu_i$ are computed based on the conditional structure implied by $\tilde{v}(i)$, as in \citep{datta2016hierarchical}.

While max-min ordering often yields the best results in Vecchia approximations \citep{guinness2018permutation}, applying it directly in this censored-data context is problematic, since computing conditional expectations requires knowledge of the unobserved censored values. To circumvent this issue, we propose an ordering strategy that first places all non-censored observations—ordered via the max-min criterion—followed by the censored ones in arbitrary order. Conditioning is then restricted to the nearest $M$ preceding non-censored observations.

\begin{figure}[ht]
    \centering
    \includegraphics[width=\linewidth]{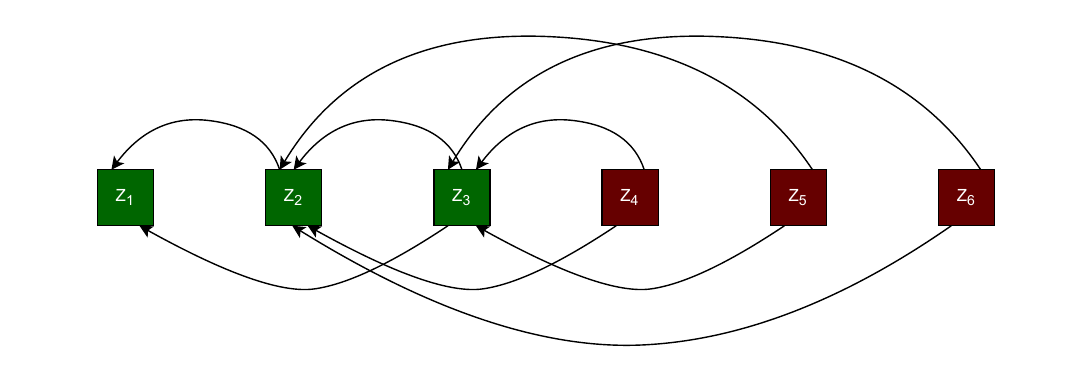}
    \caption{Toy example illustrating the latent-free Vecchia approximation with 3 observed and 3 censored values. Conditioning is restricted to the two closest preceding non-censored observations.}
    \label{fig:mon_image}
\end{figure}

A toy example is provided in Figure~\ref{fig:mon_image}, where we consider $n=6$ observations (3 observed and 3 censored), with $M=2$. The non-censored observations are ordered first and each observation conditions on its $M$ closest predecessors that are also non-censored.

This construction enables evaluation of the (approximated) data likelihood in a single pass over the data, given parameters $\alpha$, $\tau^2$, and $\Theta$, without the need for a latent process and with matrix inversions of size at most $M \times M$. This results in a highly scalable inference scheme for censored Bayesian SVC models.

\subsection{Predictions}
Following the MCMC sampling procedures described previously, we obtain $L$ posterior samples ${(\alpha, \Theta, \tau^{2}, w)}$ for Full Latent model and Latent Vecchia, and ${(\alpha, \Theta, \tau^{2})}$ for Latent-Free Vecchia model.

We first introduce the notation for prediction locations, and then give for each of the three models :
\begin{enumerate}
    \item The formula for the posterior mean for each vector of the posterior distribution of the parameters.
    \item The formula for conditional simulation at unsampled locations. 
\end{enumerate}

Let
\begin{itemize}
    \item $X_o=\{x_i:i\in o\}$ be the covariates at the $n_O$ observed sites.
    \item $X_p=\{x_j:j\in p\}$ be the covariates at the $n_P$ new (unsampled) sites.
    \item $w_o=(w_i:i\in o)$ the latent Gaussian vector at observed sites.
    \item $w_p=(w_j:j\in p)$ the latent Gaussian vector at prediction sites.
\end{itemize}
The joint distribution of the observations and the predictions remains a Gaussian process, with a covariance matrix denoted by $K$:

$$
\begin{pmatrix}w_o\\w_p\end{pmatrix}
\sim N\Bigl(\begin{pmatrix}X_o\alpha\\X_p\alpha\end{pmatrix},\,
K=\begin{pmatrix}K_{oo}&K_{op}\\K_{po}&K_{pp}\end{pmatrix}\Bigr)\,. 
$$\\

Let $Q$ be the inverse of $K$,i.e the precision matrix, and $U$ its Cholesky decomposition $K^{-1}=Q=UU^t$.
$U$ is therefore a triangular matrix 
$$
U=\begin{pmatrix}U_{oo}&U_{op}\\0&U_{pp}\end{pmatrix}\,. 
$$

\subsubsection{Full‐latent Gaussian‐process}

Because we sampled the latent vector $w_o$, standard GP conditioning applies:

The conditional mean :

$$
\mu_p
= X_p\alpha -Q_{pp}^{-1}\,Q_{po}\,(w_o-X_o\alpha).
$$

Conditional simulations can be obtained using :

$$
\operatorname{Var}(w_p\mid w_o)
= Q_{pp}^{-1},
\quad
w_p^{\rm sim}
= \mu_p + U_{pp}^{-1}\,\varepsilon. \quad \varepsilon \sim N(0,I)
$$

\subsubsection{ Vecchia‐latent}
Again we have a sample of $w_o$, under the constraint that the prediction points are ordered after the observation points, we can use the results of \citep{katzfuss2020vecchia} to predict using the Vecchia approximation :

$$
\mu_p
= X_p\alpha -(U_{pp}^{\;T})^{-1}\,U_{op}^{\;T}\;(w_o-X_o\alpha),
$$

The prediction tasks can be carried out solely based on \( U_{\bullet p} \), which is the sub-matrix formed by the last \( n_P \) columns of \( U \) corresponding to \( w_p \). That is, the first \( n_O \) columns of \( U \) corresponding to \( w_o \)  would then not be required for prediction, resulting in a prediction complexity that depends on \( n_P \), not on \( n = n_O + n_P \). This computational simplification comes at the price of some loss of accuracy.

The non zeros entries of  \( U_{\bullet p} \) can be directly calculated knowing the covariance parameters and the number of neighbors M by following the algorithm available in \citep{katzfuss2020vecchia}.

And so a conditional draw is simply :

$$
w_p^{\rm sim}
= \mu_p + U_{pp}^{-1}\,\varepsilon,
\quad \varepsilon\sim N(0,I).
$$

\subsubsection{ Vecchia-latent‐free }

Here $w_o$ is not sampled. Prediction at unsampled locations is carried out in two steps:
in the first step, we predict the response at censored locations using the non-censored observations and adjust these predictions with the Mills ratio \citep{heckman1979sample}. In the second step, we predict at the truly unsampled locations using both the non-censored observations and the newly recovered values at the censored locations.

\textbf{Step 1: predict at censored locations :
}
After ordering $Z_o=(Z_{o\setminus c},\,Z_c)$ with $Z_{o\setminus c}$ ordered via max-min ordering, and $Z_c$ condition on the M nearest $Z_{o\setminus c}$.

We can build the Vecchia Cholesky factor $U^{(1)}$ of the precision matrix  of $(Z_{o\setminus c},Z_c)$. Next, we can compute

     $$
     \mu_c
     = X_c\alpha - \bigl(U^{(1)}_{cc}{}^{T}\bigr)^{-1}\,
       \bigl(U^{(1)}_{(o\setminus c),\,c}{}^{T}\bigr)\,
       (Z_{o\setminus c}-X_{o\setminus c}\alpha),
     \quad
     K_{c}
     = \bigl(U^{(1)}_{cc}{}^{T} U^{(1)}_{cc}\bigr)^{-1}.
     $$
To compute $\mu_c^*=E[Z_c|Z_{o\setminus c},w_c<L]$ we adjust using the Mills ratio.

The Mills ratio adjustment arises naturally from truncated normal theory. For a normal variable $Z \sim \mathcal{N}(\mu, \sigma^2)$ censored at $L$, the conditional expectation is $E[Z | Z \leq L] = \mu - \sigma \lambda((L - \mu)/\sigma)$, where $\lambda$ is the inverse Mills ratio \citep{heckman1979sample}. The Vecchia approximation induces conditional independence among observations given their conditioning sets, allowing this univariate result to be applied componentwise to obtain the exact censored conditional mean $\mu_c^*$ under the approximated model.

\[
\mu_c^*
=
\mu_c
-
\diag(K_c^{1/2})
\lambda\left(
\frac{L - \mu_c}{\diag(K_c^{1/2})}
\right)
\]
where \(\lambda(a) = \frac{\phi(a)}{\Phi(a)}\) is the inverse Mills ratio,
with \(\phi\) the standard normal density and \(\Phi\) the standard normal Cumulative distribution function, applied component-wise.\\

\textbf{Stage 2: Predict at new sites :}

  We begin by augmenting the observed data through $\widetilde Z_o=(\,Z_{o\setminus c},\,\mu_c^*\,)$.
The augmented vector $\widetilde Z_o$ is then ordered, prediction last, using max-min approach, from which we construct the Vecchia Cholesky decomposition $U^{(2)}$.
   Following the methodology established previously :
     $\;U^{(2)}=\begin{pmatrix}U^{(2)}_{oo}&U^{(2)}_{op}\\0&U^{(2)}_{pp}\end{pmatrix}$,
     and set
     $$
     \mu_p
     = X_p\alpha-\bigl(U^{(2)}_{pp}{}^{T}\bigr)^{-1}
       \bigl(U^{(2)}_{op}{}^{T}\bigr)\,
       (\widetilde Z_o-X_o\alpha),
         $$
     with simulation
     $  w_p^{\rm sim}   = \mu_p + U^{(2)\,-1}_{pp}\,\varepsilon.$\\

\textbf{Summary}
All three models yield a posterior mean $\mu_p$ and conditional simulations for $w_p$.  The "Full-latent" model uses the exact dense‐GP formulas, "Latent Vecchia" uses a single Vecchia decomposition of $(w_o,w_p)$, and "Latent-Free Vecchia" uses two Vecchia steps with a Mills‐ratio adjustment to handle the censored observations.

\section{Validation and Performance Evaluation}
\label{sec:validation}

This section presents a comprehensive validation of the proposed latent-free Vecchia approximation for Bayesian spatially varying coefficients (SVC) models with censored data. We address two primary objectives: (1) evaluating the accuracy of the approximated likelihood computation, and (2) comparing the performance of three Bayesian inference approaches across different censoring scenarios and computational settings.

\subsection{Likelihood Approximation Accuracy}
\label{subsec:likelihood_accuracy}

To assess the quality of our proposed likelihood approximation, we conduct a large-scale simulation study comparing the true likelihood with the Vecchia-approximated likelihood across various censoring levels and conditioning set sizes.

\subsubsection{Simulation Setup}

We simulate 10,000 independent datasets, each containing $n = 200$ spatial observations within a unit square. Each dataset follows the SVC model with $p = 2$ predictors, using the following parameter configuration:
\begin{itemize}
    \item Fixed effects: $\alpha = (-5, 10)^\top$
    \item Spatial range parameters: $\phi = (40, 15)^\top$  
    \item Variance parameters: $\sigma^2 = (15, 30)^\top$
    \item Nugget variance: $\tau^2 = 0.1$
\end{itemize}

For each simulated dataset, we compute both the true likelihood using Genz's algorithm \citep{genz1992} for multivariate normal integrals and the proposed Vecchia approximation with conditioning set sizes $M \in \{10, 30, 50\}$. We consider five censoring scenarios: 0\%, 5\%, 25\%, 50\%, and 75\% of observations falling below the detection threshold.

\subsubsection{Results}

\begin{figure}[ht]
    \centering
    \includegraphics[width=\linewidth]{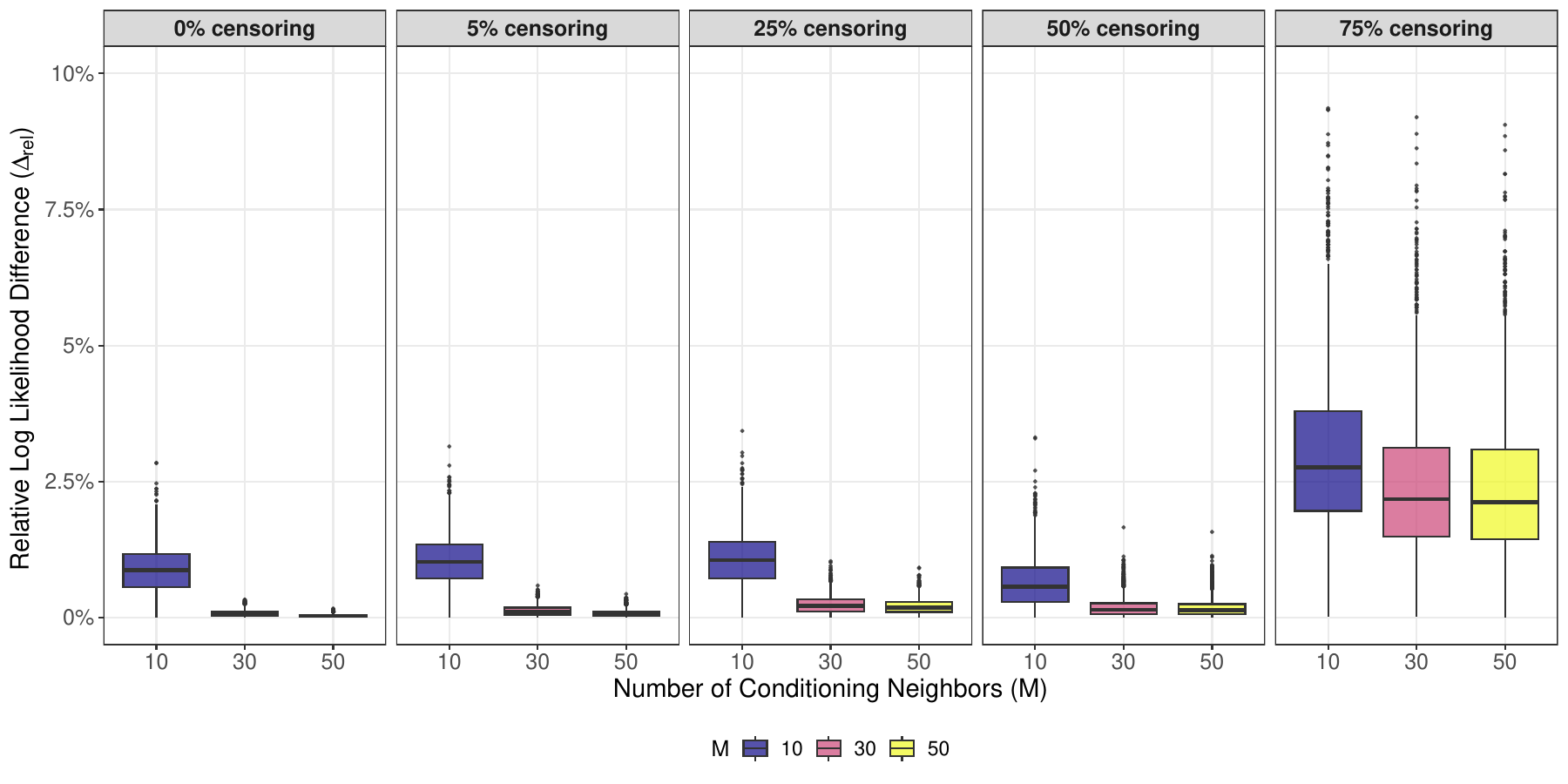}
    \caption{Relative likelihood approximation error across different conditioning set sizes ($M$) and censoring levels. The approximation demonstrates high accuracy for censoring levels below 75\%, with diminishing returns beyond $M = 30$.}
    \label{fig:likelihood_error}
\end{figure}

Figure~\ref{fig:likelihood_error} presents the relative likelihood error across different values of $M$ and censoring levels, defined as :
\[
\Delta_{\text{rel}} = \frac{|\log \tilde{p}(Z) - \log p(Z)|}{|\log p(Z)|} \times 100\%
\]

The results demonstrate that the Vecchia approximation maintains high accuracy across most practical scenarios, with particularly strong performance for censoring levels below 75\%. 

The approximation error remains consistently low (typically below 1\%) for moderate censoring levels (50\%), indicating that the proposed method provides a reliable alternative to exact likelihood computation. As expected, the approximation quality improves with larger conditioning sets, though the gains diminish beyond $M = 30$ for most censoring scenarios.

\begin{figure}[ht]
    \centering
    \includegraphics[width=\linewidth]{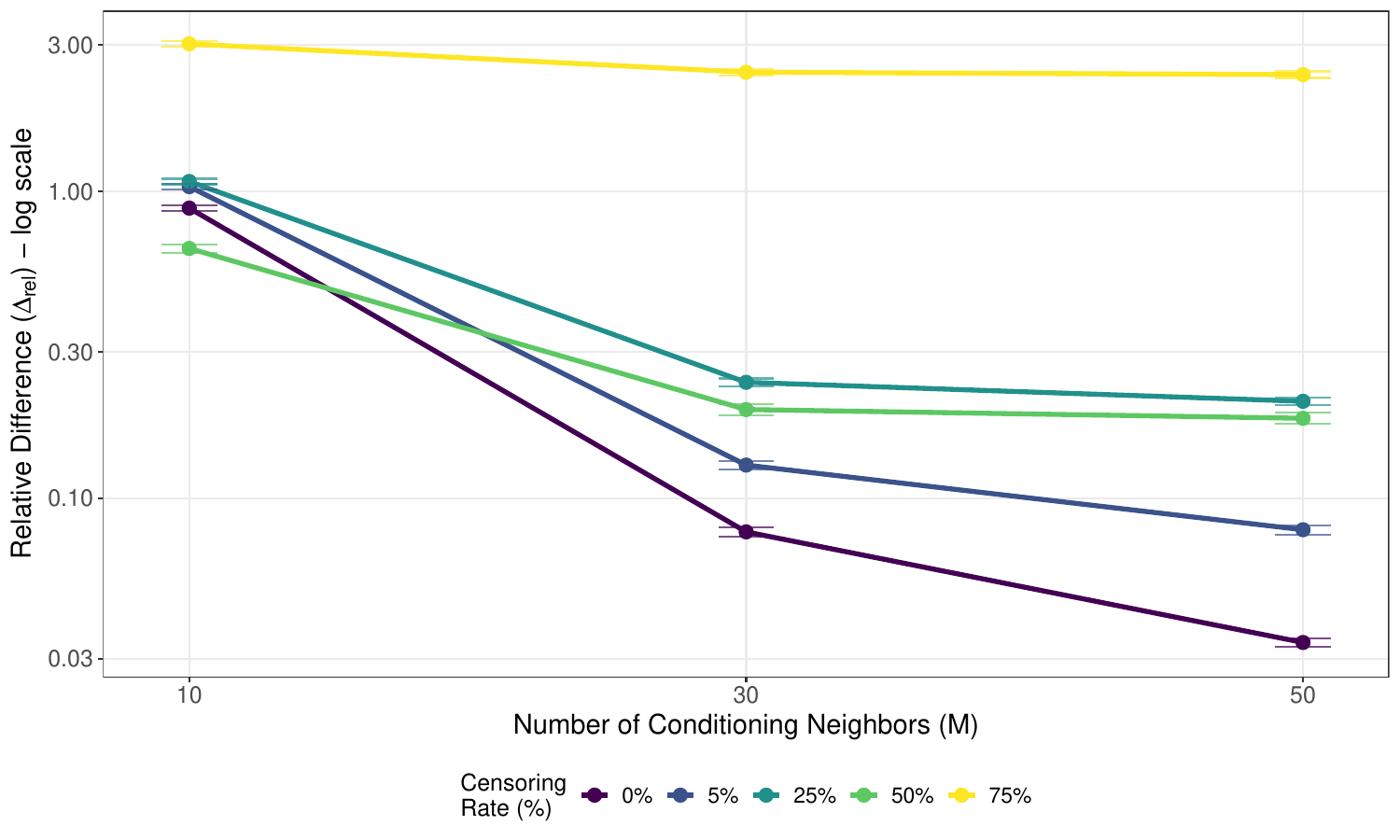}
    \caption{Evolution of likelihood approximation accuracy as a function of conditioning set size $M$ for different censoring levels. Higher censoring rates show greater improvement with increased $M$.}
    \label{fig:approximation_evolution}
\end{figure}

Figure~\ref{fig:approximation_evolution} illustrates how the approximation quality evolves as a function of $M$ for different censoring levels. This analysis reveals that lower censoring levels benefit more substantially from increased conditioning set sizes.

\subsection{Bayesian Inference Comparison}
\label{subsec:bayesian_comparison}

We compare three Bayesian inference approaches for SVC models with censored data:
\begin{enumerate}
    \item \textbf{Full Latent}: The complete latent formulation from Equation~\eqref{eq:model_priors_latent_censored}
    \item \textbf{Latent Vecchia}: Vecchia approximation applied to the latent field $w$
    \item \textbf{Latent-Free Vecchia}: Our proposed method eliminating the latent field entirely
\end{enumerate}

\subsubsection{Simulation Design}

Using a single representative dataset of $n = 200$ observations with the same parameter configuration as above, we implement MCMC sampling for each method across multiple scenarios:
\begin{itemize}
    \item Conditioning set sizes: $M \in \{10, 30, 50\}$ (for Vecchia-based methods)
    \item Censoring levels: 0\%, 5\%, 25\%, 50\%, and 75\%
    \item Chain length: 6 chains of 10,000 iterations with 5,000 burn-in
\end{itemize}

All computations were performed on a Dell Latitude 5520 laptop equipped with an Intel Core i5-1145G7 processor (8 cores, up to 4.2 GHz) and 16 GB RAM, running Ubuntu 22.04 LTS. MCMC sampling was implemented using Stan version 2.36.0 \citep{Stan} with parallel chain execution.

\subsubsection{Parameter Estimation Accuracy}

\begin{figure}[ht]
    \centering
    \includegraphics[width=\linewidth]{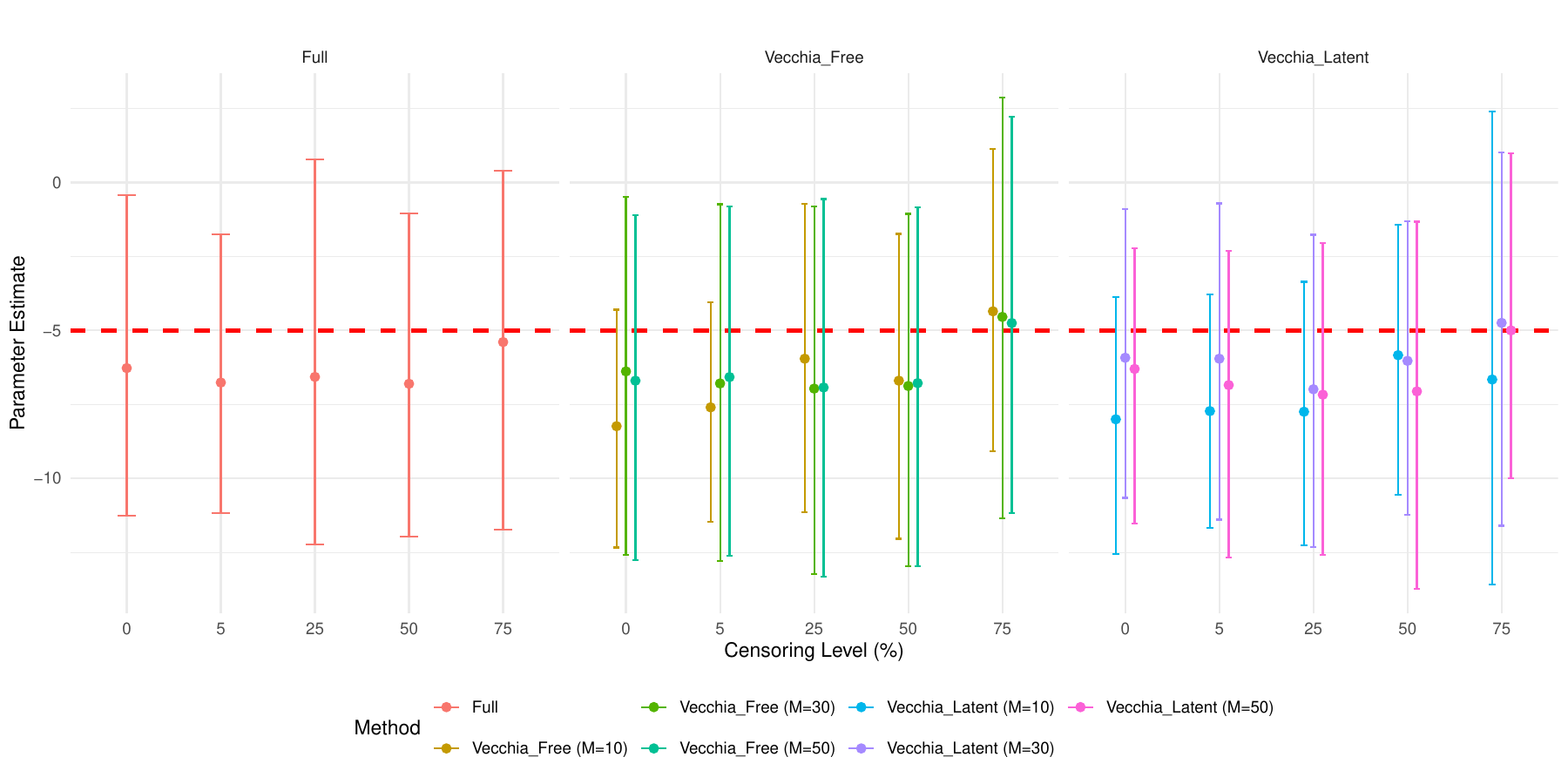}
    \caption{Posterior estimates (mean and 95\% credible intervals) for parameter $\alpha_1$ across three Bayesian methods, different conditioning set sizes, and censoring levels. The remarkable consistency between methods validates the proposed approximation.}
    \label{fig:parameter_estimation}
\end{figure}

Figure~\ref{fig:parameter_estimation} displays the posterior estimates for the first fixed effect parameter $\alpha_1$ across all methods, conditioning set sizes, and censoring levels. The results demonstrate remarkable consistency between the three approaches, with credible intervals showing substantial overlap across all scenarios.

This consistency validates our theoretical expectation that the latent-free formulation preserves the inferential properties of the full latent approach while offering computational advantages. Even under high censoring (75\%), all methods converge to similar posterior distributions, indicating robustness of the approximation scheme.

\subsubsection{Computational Efficiency}

\begin{figure}[ht]
    \centering
    \includegraphics[width=\linewidth]{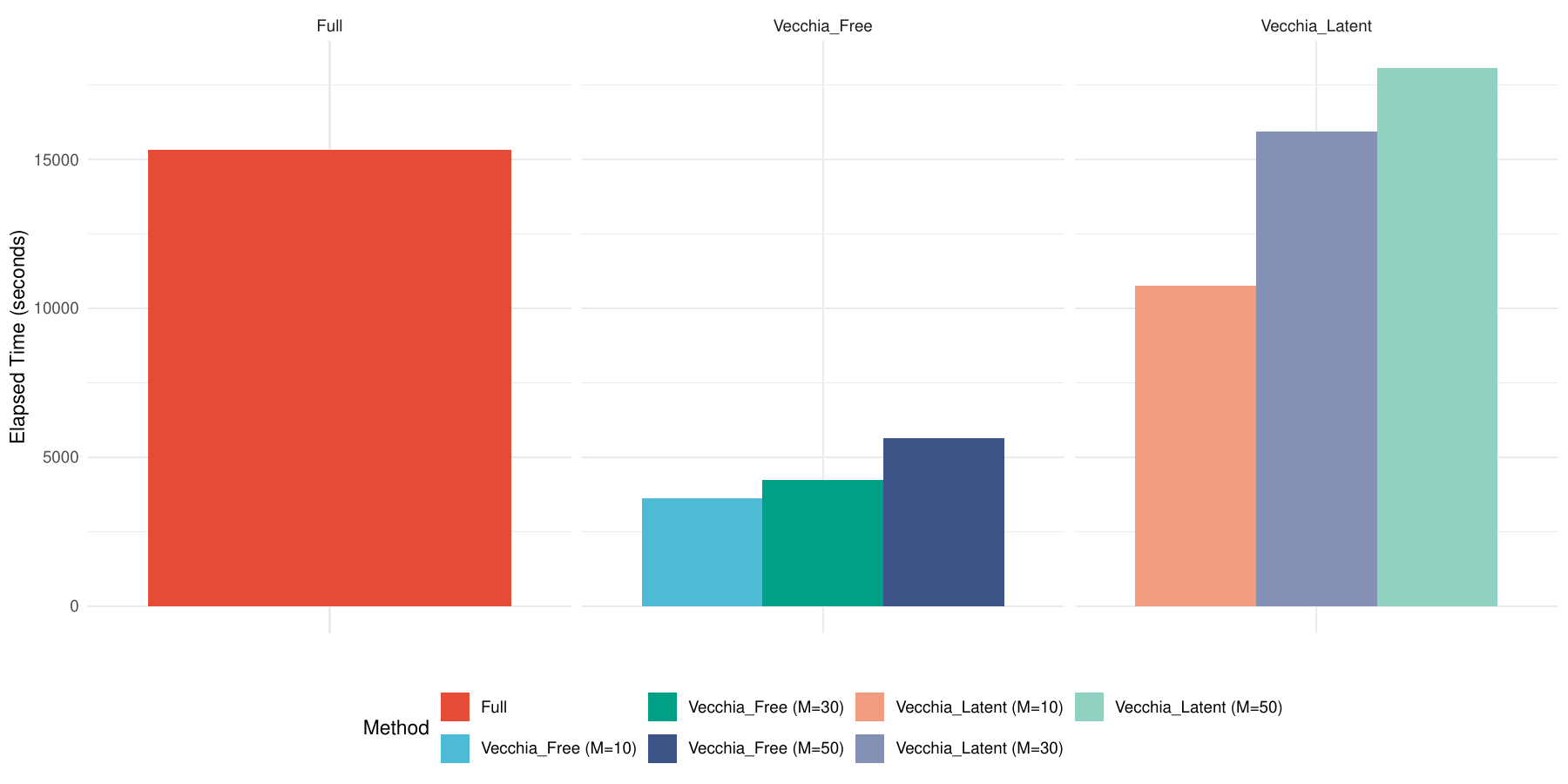}
    \caption{Computational time comparison across the three Bayesian methods for different conditioning set sizes. The latent-free Vecchia approach shows substantial speedups, with advantages expected to increase for larger datasets.}
    \label{fig:computation_time}
\end{figure}

Figure~\ref{fig:computation_time} presents the computational time comparison between the three methods for different values of $M$. The latent-free Vecchia approach demonstrates substantial computational advantages, achieving significant speedups compared to the full latent approach.

Importantly, these timing differences represent a conservative assessment, as they are measured on relatively small datasets ($n = 200$). The computational advantages of the latent-free method are expected to become even more pronounced for larger spatial datasets, where the elimination of the $n$-dimensional latent field $w$ from the parameter space yields increasingly significant efficiency gains.

\subsubsection{Predictive Performance Comparison}

To evaluate the predictive capabilities of the three bayesian approaches, we assess their performance on a holdout test dataset consisting of 800 spatial locations. We compute out-of-sample predictions using 30,000 posterior samples from each fitted model and evaluate two distinct aspects of predictive performance using complementary metrics.\\
For point prediction assessment, we use Root Mean Square Error (RMSE) to evaluate the quality of the conditional mean predictions. The conditional mean at each test location is computed by averaging over all 30,000 krigings, providing a single point estimate that represents the expected value of the response variable given the observed data and model parameters.\\
For distributional prediction assessment, we employ the Continuous Ranked Probability Score (CRPS) to validate the quality of the conditional simulations. Rather than collapsing to a point estimate, this evaluation utilizes the full set of 30,000 posterior simulations to construct an empirical distribution of the response variable at each test location. The CRPS then measures how well this entire predictive distribution matches the observed test values, providing a comprehensive assessment of the model's ability to quantify prediction uncertainty. Lower CRPS values indicate better probabilistic forecasting accuracy.\\

\begin{figure}[ht]
\centering
\includegraphics[width=\linewidth]{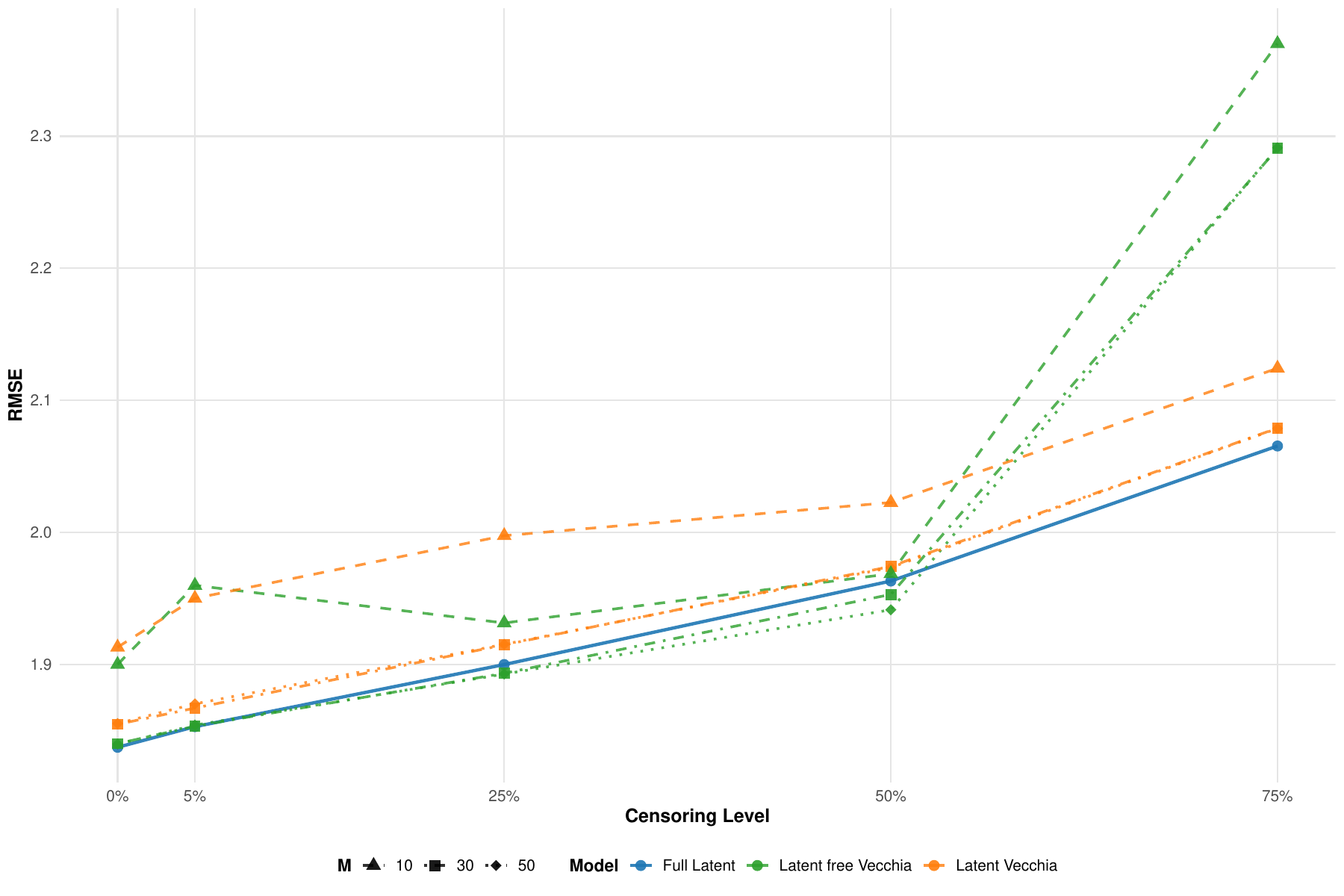}
\caption{RMSE performance across censoring scenarios for the three Bayesian methods.}
\label{fig:rmse_comparison}
\end{figure}

Figure~\ref{fig:rmse_comparison} presents the RMSE performance for the  three models across all censoring scenarios and conditioning set sizes. The results reveal several important patterns regarding the impact of censoring and approximation parameters on prediction accuracy.

\begin{figure}[ht]
\centering
\includegraphics[width=\linewidth]{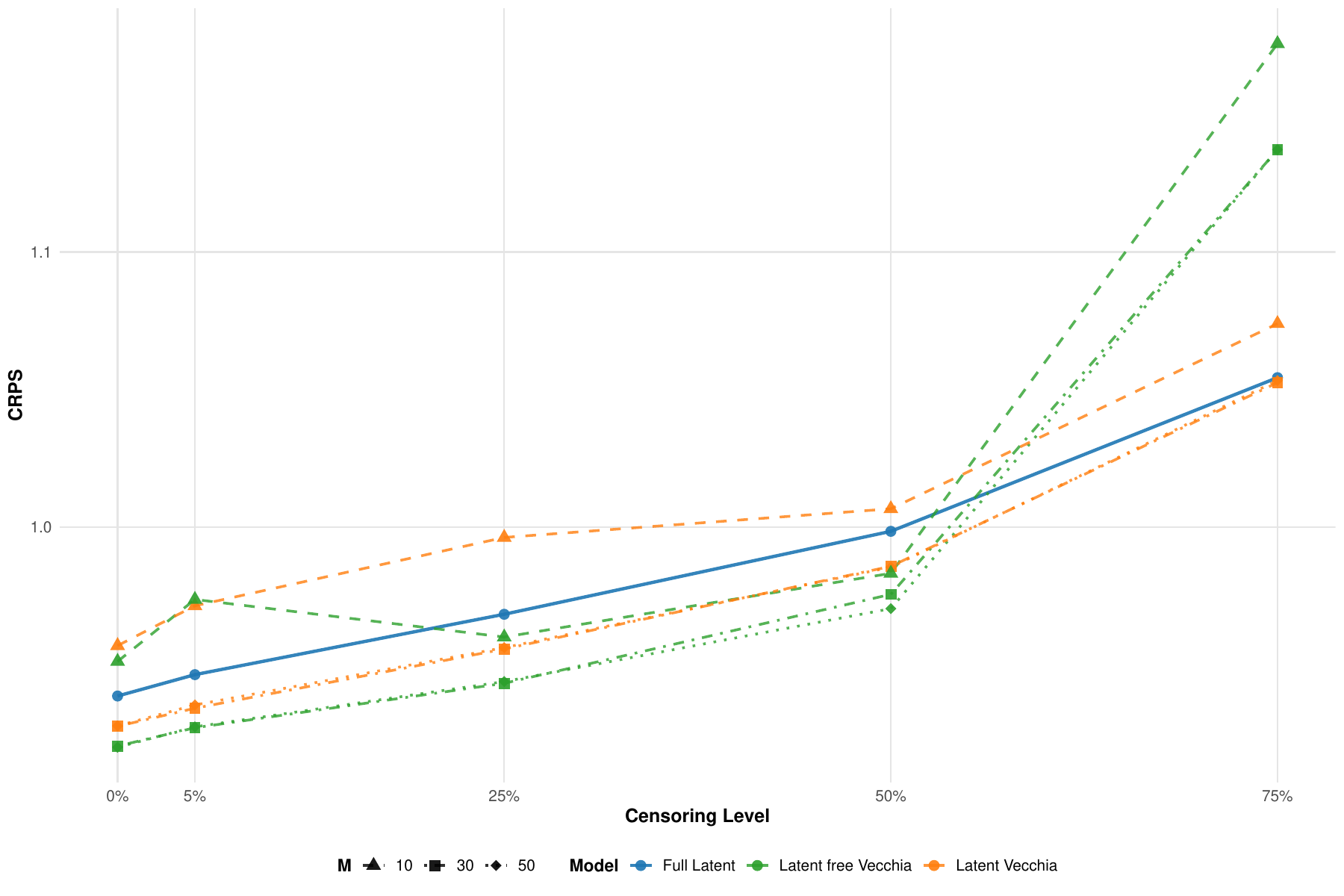}
\caption{CRPS performance across censoring scenarios for the three Bayesian methods.}
\label{fig:crps_comparison}
\end{figure}

Figure~\ref{fig:crps_comparison} shows the corresponding CRPS results, which provide insight into the quality of the entire predictive distribution rather than just point estimates. The CRPS results largely mirror the RMSE patterns, confirming that the methods' relative performances are consistent across both point and distributional prediction metrics.

Regarding the effect of censoring, as expected, all methods exhibit deteriorating prediction performance as the censoring level increases from 0\% to 75\%. This degradation reflects the fundamental challenge of making predictions when substantial portions of the training data are censored. However, the rate of deterioration varies across the methods, with model 3 performances deteriorating more in the 75\% censoring scenario, as observed previously for the likelihood approximation accuracy in figure \ref{fig:approximation_evolution}.

As for the impact of the conditioning set size, the results reveal that for the Vecchia-based approaches (Latent Vecchia and Latent-Free models), increasing the conditioning set size $M$ from 10 to 50 leads to consistent improvements in prediction accuracy across all censoring scenarios. This enhancement becomes particularly evident under higher censoring levels, where the additional conditioning information proves more valuable for achieving accurate approximations.

Interestingly, we observe that Latent-Free model can outperform the Full-latent model under certain censoring scenarios. While this might initially appear counterintuitive, the explanation lies in the nature of the prediction task, which incorporates the quality of parameter estimation. Since Latent-Free model does not rely on a latent field, it converges considerably faster than the Full-latent model. This faster convergence allows the model to provide a more accurate representation of the posterior parameter distribution within the same number of iterations.
When we compare the models using an equal number of MCMC iterations, Latent-Free model emerges as both faster and more effective in terms of predictive performance, particularly when the censoring threshold remains below 50\% and the conditioning set is relatively large (M=30 or 50).

Taken together, our estimation and prediction results demonstrate that Free-Latent Vecchia model offers a practical and efficient framework for modeling Bayesian spatially varying coefficients in the presence of censoring. Its computational efficiency and robust performance under moderate to high censoring make it a compelling choice for applied spatial analysis.

\subsection{Summary}

The validation study demonstrates two key achievements of our proposed methodology:

\textbf{Likelihood Approximation Quality:} The relative error analysis confirms that the Vecchia approximation to the censored-data likelihood maintains high accuracy across practical censoring scenarios. The approximation performs exceptionally well for censoring levels up to 50\%, with acceptable performance even at 75\% censoring when appropriate conditioning set sizes are employed.

\textbf{Computational Scalability:} The latent-free formulation successfully eliminates the primary computational bottlenecks associated with high-dimensional latent fields while preserving inferential accuracy. The dramatic reduction in computational time, combined with the maintained parameter estimation quality, makes this approach particularly attractive for large-scale spatial applications.

The consistency of parameter estimates across all three methods provides strong evidence that the proposed approximations do not introduce substantial bias, while the computational gains position this methodology as a practical solution for analyzing large censored spatial datasets with spatially varying coefficients.

These results establish the foundation for applying Bayesian SVC models to real-world problems where both computational efficiency and statistical rigor are essential, particularly in environmental and epidemiological applications where censored observations are commonplace.

\section{Case Study: Hydrocarbon Contamination in Toulouse Metropolis}
This section demonstrates the practical applicability of our proposed latent-free Vecchia approximation for Bayesian spatially varying coefficients models with censored data. The case study analyze hydrocarbon contamination data from the Toulouse Metropolis, France. It represents a typical environmental monitoring scenario where detection limits create substantial censoring challenges, and spatial heterogeneity necessitates flexible modeling approaches.

\subsection{Data Description and Study Area}

The dataset comprises soil samples collected across the Toulouse metropolitan area, with measurements of hydrocarbon concentrations (C10-C40) at various locations within the study region. The spatial domain covers approximately 31.25 km² (555.46–586.71 km longitude, 6268.91–6294.11 km latitude) in the Lambert Conformal Conic projection system. The irregular sampling design reflects the practical constraints of environmental monitoring, with higher sampling density near urban areas and sparser coverage in peripheral zones.

Hydrocarbon contamination in urban environments typically originates from multiple sources including fuel storage facilities, industrial activities, vehicular emissions, and historical spills. The C10-C40 fraction represents medium to heavy petroleum hydrocarbons, which are particularly relevant for soil contamination assessment due to their persistence and potential environmental impacts.

For confidentiality reasons, no numerical values of hydrocarbon concentrations are reported.

Based on established environmental and urban development theory, we consider six spatially-referenced predictors that potentially influence hydrocarbon distribution patterns:

\textbf{Topographic Factor:}
\begin{itemize}
    \item \textbf{Altitude (ALT):} Elevation above sea level (meters), ranging from 107.2 to 281 meters. Topography influences drainage patterns, erosion processes, and pollutant transport mechanisms, with lower elevations potentially accumulating hydrocarbons through gravitational flow and surface runoff.
\end{itemize}

\textbf{Distance-Based Proximity Measures:}
\begin{itemize}
    \item \textbf{dist\_basias:} Distance to potentially contaminated sites (meters), based on the BASIAS database of potentially polluting activities, including fuel stations, industrial facilities, and historical contamination sources.
    \item \textbf{dist\_basol:} Distance to confirmed polluted or remediated sites (meters), from the BASOL database of contaminated sites requiring monitoring, many of which involve hydrocarbon contamination.
    \item \textbf{dist\_center:} Distance to the city center (meters), capturing urban development patterns and historical contamination sources, with city centers often having higher concentrations of potential hydrocarbon sources.
    \item \textbf{dist\_water:} Distance to rivers and water bodies (meters), reflecting hydrological transport pathways and potential dilution effects for hydrocarbon contamination.
    \item \textbf{dist\_road:} Distance to the nearest road (meters), representing traffic-related hydrocarbon contamination sources including fuel spills, vehicle emissions, and road runoff.
\end{itemize}

An intercept term is included to account for the baseline level, resulting in a total of seven covariates.

The dataset exhibits several characteristics that makes it particularly suitable for demonstrating our methodology:

\textbf{Left-Censoring Issues:} Approximately 50\% of observations fall below the detection limit, creating substantial censoring that must be appropriately handled to avoid biased parameter estimates and predictions. Importantly, the censored observations show no apparent spatial clustering pattern.

\textbf{Computational Scale:} With $n = 136$ observations and $p = 7$ predictors, this dataset represents a moderately-sized spatial problem. While this scale remains computationally feasible, implementing a full Bayesian inference with latent spatial fields would be prohibitively expensive due to the extensive MCMC sampling required. Moreover, the prediction grid contains 10,000 locations, making it computationally infeasible to perform kriging at each MCMC iteration without approximation. This computational constraint suggests the use of our proposed model without an explicit latent field specification, offering a more tractable approach to spatial modeling at this scale

\subsection{Model Specification and implementation details}

We apply our latent-free Vecchia SVC model to the log-transformed hydrocarbon concentration data, using the following specifications:

For each environmental covariate $j \in \{1,\ldots,7\}$, we model the spatially varying coefficients as:
\begin{equation}
\beta_j(s) = \alpha_j + \eta_j(s)
\end{equation}

where $\alpha_j$ represents the global mean effect and $\eta_j(s)$ is a zero-mean Gaussian process with exponential covariance function :
\begin{equation}
\text{Cov}(\eta_j(s), \eta_j(s')) = \sigma^2_j \exp\left(-\phi_j \|s - s'\|\right)
\end{equation}
where $\sigma^2_j$ is the variance parameter, $\phi_j$ is the decay (or inverse range) parameter, and $\|s - s'\|$ denotes the Euclidean distance between spatial locations $s$ and $s'$.

We employ weakly informative prior distributions: Gaussian priors for the global 
mean effects $\alpha_j$, inverse-gamma priors for the variance parameters 
$\sigma^2_j$ and $\tau^2$, and gamma priors for the spatial decay parameters 
$\phi_j$. These choices reflect standard practice in spatial statistics while allowing the data to dominate posterior inference. Specific hyperparameter values were chosen to be weakly informative given the scale of the study area and the anticipated magnitude of covariate effects.


We implement the latent-free Vecchia approximation with conditioning set size $M = 40$. The spatial ordering follows our proposed strategy: non-censored observations are ordered using the max-min criterion, followed by censored observations, with conditioning restricted to the $M$ nearest preceding non-censored neighbors.

To generate spatial predictions across the study area, we construct a regular prediction grid containing approximately 10,000 locations, with covariate values extracted at each grid point using the environmental predictors described in the previous subsection. MCMC sampling is implemented in Stan with 4 parallel chains, each running for 10,000 iterations (including 5,000 burn-in samples), resulting in 
20,000 posterior samples. Consequently, we obtain 20,000 posterior predictive realizations at each of the 10,000 grid locations, enabling comprehensive uncertainty quantification for the spatial predictions.

\subsection{Results and Spatial Patterns}

The Bayesian spatially varying coefficients model with latent-free Vecchia approximation successfully converged after 10,000 MCMC iterations across 4 parallel chains. All parameters exhibited satisfactory mixing properties and convergence diagnostics (all $\hat{R} < 1.01$).

\begin{figure}[H]
\centering
\includegraphics[width=\linewidth]{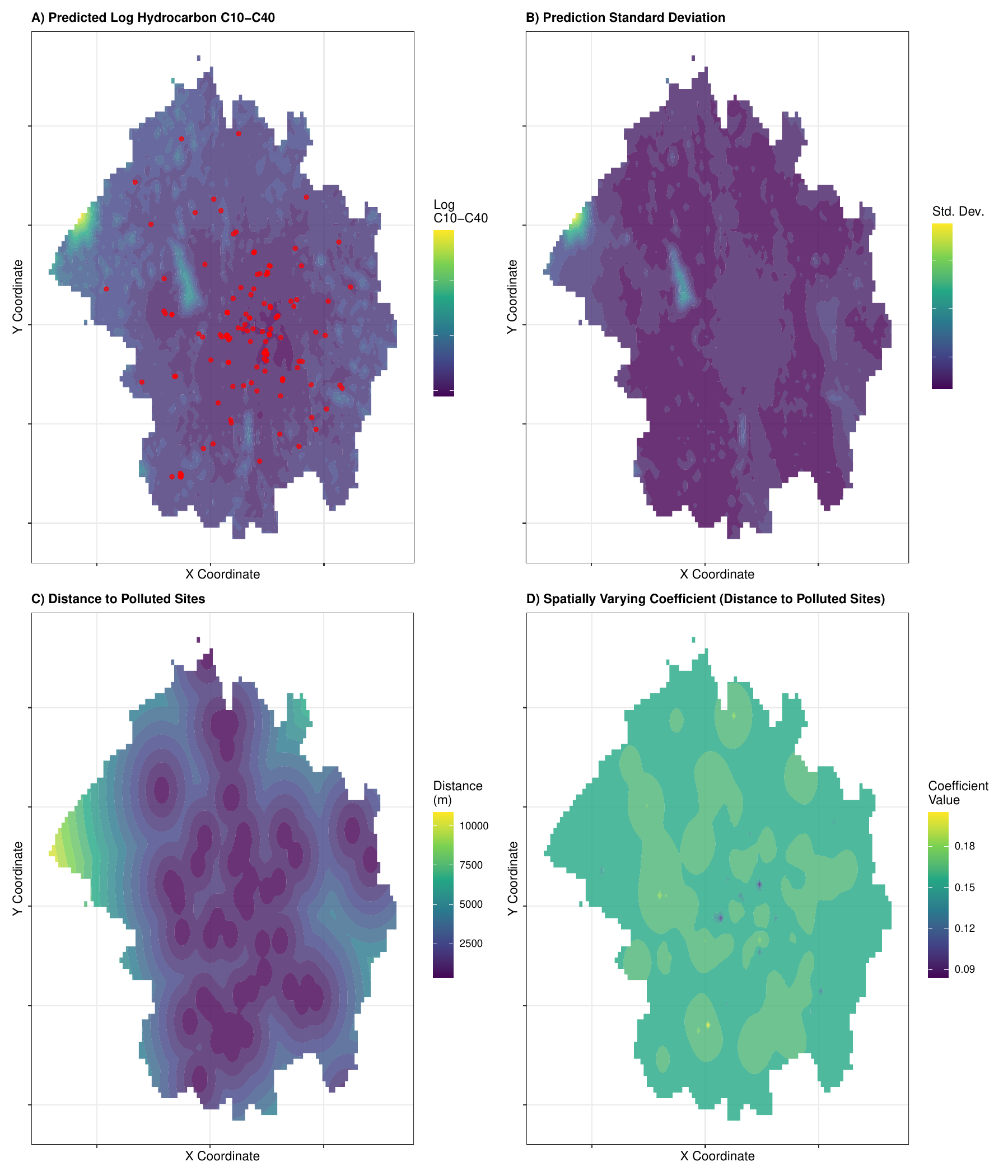}
\caption{Spatial analysis of hydrocarbon contamination in Toulouse Metropolis. (A) Posterior mean predictions of log-transformed C10-C40 concentrations with observed sample locations (red points). (B) Prediction uncertainty (posterior standard deviation). (C) Distance to confirmed polluted sites (BASOL database). (D) Spatially varying coefficient for distance to polluted sites, showing heterogeneous effects across the study area}
\label{fig:spatial_analysis}
\end{figure}

Figure~\ref{fig:spatial_analysis} presents four complementary perspectives on the spatial analysis of hydrocarbon contamination. \\

Panel A displays the posterior mean predictions of log-transformed hydrocarbon concentrations (C10-C40) across the study area. The spatial pattern reveals substantial heterogeneity in contamination levels, with higher concentrations observed in the vicinity of Toulouse Airport and across the northwestern part of the region. The red points indicate the locations of observed soil samples, demonstrating the irregular sampling design with denser coverage in certain urban sectors.

Panel B illustrates the prediction uncertainty through the standard deviation of the posterior predictive distribution. The uncertainty map reveals an intuitive pattern: prediction variance is lowest near observed sample locations and increases progressively with distance from sampling points. Notably, regions with higher predicted contamination levels coincide with areas of high uncertainty, These high-value predictions in data-sparse regions reflect covariate-driven extrapolation, where the model identifies relationships between contamination and environmental predictors. The spatial distribution of uncertainty provides valuable guidance for adaptive sampling strategies, identifying areas where additional measurements would most effectively reduce predictive uncertainty.

Panel C presents one of the key environmental covariates: distance to confirmed polluted sites (BASOL database). The spatial pattern reveals that certain zones of the metropolis are in close proximity to known contamination sources, while peripheral areas are more distant from documented polluted sites. 

Panel D displays the spatially varying coefficient for this covariate (mean-centered), revealing substantial spatial heterogeneity in how proximity to documented sites influences hydrocarbon concentrations. While conventional wisdom suggests proximity to known contaminated sites should correlate with higher pollution levels, the spatially varying coefficients demonstrate a more complex reality: the relationship varies considerably across the study area, with both positive and negative local effects. This heterogeneity likely reflects the interplay of multiple factors including site remediation history, incomplete database coverage of diffuse pollution sources, spatial confounding with other environmental variables, and local-scale transport processes. The flexibility of the SVC framework proves essential here, as a global coefficient would mask these important spatial variations in the contamination-distance relationship.

\section{Conclusion}
This paper introduces an efficient Bayesian framework for spatially varying coefficients models that addresses two critical challenges in environmental monitoring: computational scalability and the presence of censored observations. Our proposed latent-free Vecchia approximation eliminates the high-dimensional latent field from the parameter space while maintaining inferential accuracy, achieving substantial computational gains without sacrificing statistical rigor.

Through comprehensive validation studies, we demonstrated that the Vecchia approximation maintains high accuracy for censoring levels up to 75\%, with optimal performance when the conditioning set size is appropriately chosen. The comparative analysis across three Bayesian approaches revealed that our latent-free formulation not only reduces the computational burden but also exhibits faster MCMC convergence, making it particularly attractive for large-scale spatial applications.

The case study of hydrocarbon contamination in Toulouse Metropolis illustrates the practical value of this methodology for real-world environmental assessment. The model successfully captured spatially heterogeneous relationships between contamination levels and environmental predictors, providing interpretable insights into the complex processes governing pollutant distribution. Importantly, the spatially varying coefficients revealed that the influence of pollution sources is not uniform across the study area, highlighting the importance of flexible modeling frameworks that can accommodate spatial non-stationarity.

Our approach offers several advantages for practitioners: (1) it scales efficiently to moderately large datasets with multiple covariates, (2) it naturally handles censored observations without introducing additional latent variables, (3) it provides full uncertainty quantification through Bayesian inference, and (4) it maintains model interpretability through spatially explicit coefficient estimates. These features make the methodology particularly well-suited for environmental risk assessment, where both statistical rigor and computational feasibility are essential.

Future research directions include extending the framework to multivariate contamination modeling, incorporating temporal dynamics for monitoring data, and developing adaptive sampling strategies guided by predictive uncertainty maps. Additionally, investigating alternative spatial ordering schemes and conditioning strategies may further improve the approximation quality under extreme censoring scenarios. The proposed methodology establishes a foundation for scalable Bayesian spatial analysis in settings where data limitations and computational constraints have traditionally hindered the application of flexible modeling approaches.

\bigskip 

\textbf{Acknowledgements:} We acknowledge financial funding by ANR-HOUSES (grant number: ANR-22-CE56-0006).

\bibliographystyle{elsarticle-harv} 
\bibliography{bibifile}



\end{document}